\documentclass[aps,prd,showpacs,nofootinbib,twocolumn]{revtex4}
\usepackage{latexsym}
\usepackage{amsmath,amsfonts}
\usepackage{amsbsy}
\usepackage{mathrsfs}
\usepackage{graphicx,calc,epsfig}
\def\ut#1{\rlap{\lower1ex\hbox{$\sim$}}#1{}}

\newcommand{\C}{\mathbb{C}}
\newcommand{\R}{\mathbb{R}}
\newcommand{\be}{\nopagebreak[3]\begin{equation}}
\newcommand{\ee}{\end{equation}}
\newcommand{\ba}{\nopagebreak[3]\begin{eqnarray}}
\newcommand{\ea}{\end{eqnarray}}
\DeclareFontFamily{U}{rsfs}{}         % Formal Script            %
\DeclareFontShape{U}{rsfs}{m}{n}{<5> rsfs5 <6><7> rsfs7          %
  <8><9><10><10.95><12><14.4><17.28><20.74><24.88> rsfs10}{}     %
\DeclareMathAlphabet{\mathfs}{U}{rsfs}{m}{n}                     %
\newcommand{\mfs}[1]{\mathfs {#1}}                               %
\newcommand{\va}{\scriptscriptstyle}
\newcommand{\inter}{{\lrcorner}}
\newcommand{\van}{\scriptstyle}

\newcommand{\n}{{\nonumber}}

\newcommand{\sH}{{\mfs H}}
\newcommand{\sL}{{\mfs L}}

\newcommand{\sN}{{\mfs N}}
\newcommand{\sM}{{\mfs M}}

\newcommand{\Lie}{\sL}

\newcommand{\Vect}{\mathrm{Vect}}

\begin{document}

\title{Black hole entropy and $SU(2)$ Chern-Simons theory}

\date{\today}

\author{Jonathan Engle $^1$}

\author{Karim Noui $^2$}

\author{Alejandro Perez$^1$}

\affiliation{$^1$Centre de Physique Th\'eorique\footnote{Unit\'e Mixte
de Recherche (UMR 6207) du CNRS et des Universit\'es Aix-Marseille
I, Aix-Marseille II, et du Sud Toulon-Var; laboratoire afili\'e
\`a la FRUMAM (FR 2291)}, Campus de Luminy, 13288 Marseille,
France.}

\affiliation{$^2$
Laboratoire de Math\'ematique et Physique Th\'eorique\footnote{F\'ed\'eration Denis Poisson Orl\'eans-Tours, CNRS/UMR 6083}, 37200 Tours, France.}

\begin{abstract}

Black holes in equilibrium can be defined locally in terms of the
so-called {\em isolated horizon} boundary condition given on a null
surface representing the event horizon.  We show that this boundary
condition can be treated in a manifestly $SU(2)$ invariant manner.
Upon quantization, state counting is expressed in terms of the
dimension of Chern-Simons Hilbert spaces 
%(with level $k=a_{\va H}/(2\pi \ell_p^2\beta (1-\beta^2))$) 
on a sphere with punctures. Remarkably, when considering an ensemble of fixed horizon area $a_{\va H}$, 
the counting can be mapped to simply counting the number
of $SU(2)$ intertwiners compatible with the spins labelling the
punctures. The resulting BH entropy is proportional to $a_H$ with logarithmic corrections
$\Delta S=-\frac{3}{2} \log a_{\va H}$. Our treatment from first principles settles
previous controversies concerning the counting of states.
\end{abstract}

%\pacs{}

\maketitle

%%%%%%%%%%%%%%%%%%%%%%%%%%%%%%%%%%%%%%%%%%%%%%%%%%%%%%%%%%%%%%%%%%%%%%%%%%%%%%

%\section{Introduction}

Black holes are intriguing solutions of  general relativity
describing the physics of gravitational collapse. { These
fascinating systems---whose existence in our universe is supported
by a great amount of observational evidence---are remarkably simple.
%in that they are generally made only from the geometry of
%space-time.  
However, in the interior of the event horizon, the
predictive power of classical general relativity breaks down due to
the unavoidable appearance of un-physical divergences of the
gravitational field ({\em singularities}).} Dimensional arguments
imply that quantum effects cannot be neglected near 
{\em singularities}. In this precise sense,
 black holes (BH) provide  the most
tantalizing theoretical evidence for the need of a more fundamental
(quantum) description of the gravitational field.

Quantum effects are also important outside the horizon. Indeed the
semiclassical calculations of Hawking \cite{Hawking:1974sw} show
that BH's radiate as perfect black bodies at temperature
proportional to their surface gravity and have an entropy $S=a_{\va
H}/4\ell_p^2$, where $\ell_p^2=G \hbar/c^3$ is the Planck area. This
entropy is expected to arise from the huge number of microstates of
the underlying fundamental quantum theory describing the BH,
and therefore its computation from basic principles is an important
test of any candidate quantum theory of gravity. This letter
proposes a new and more fundamental framework for the computation of BH entropy in loop quantum
gravity (LQG) {and  establishes a precise relationship between 
$SU(2)$ Chern-Simons (CS) theory and quantum black hole physics as first explored in \cite{krasnovy}.
%\footnote{\bf The possible role of $SU(2)$ CS in describing 3d surface degrees of freedom in 
%4d quantum gravity was first proposed in \cite{smolin}}

Our treatment clarifies the description of both the classical
as well as the quantum theory of black holes in LQG making the
full picture more transparent.  We show that, in contrast with prior
results} \cite{bhe}, the gauge symmetry of LQG need not be reduced from $SU(2)$ to $U(1)$
at the horizon.  Even when the $U(1)$ reduction is perfectly viable at the classical
level, it leads to imposition of certain components of the quantum constraints only in a {\em weak} sense. Our $SU(2)$ invariant formulation---equivalent to the $U(1)$ at the classical level---avoids this issue
and allows the imposition of the constraints {\em strongly} in the Dirac
sense.
%Even when the $U(1)$ reduction is perfectly viable at the classical level, it 
%leads to difficulties in the quantum theory that were previously resolved by the imposition of 
%the quantum constraints only in a {\em weak} sense. Our $SU(2)$ invariant formulation---equivalent to the $U(1)$ at the classical level--- 
%simply avoids this issue and allows the imposition of the constraints {\em strongly} in the Dirac sense. 
This leads to a drastic simplification 
of the quantum theory in which states of a black
hole are now in one-to-one correspondence with the fundamental basic
volume excitations of LQG given by single intertwiner states. This settles
certain controversies concerning the relevant quantum numbers to be
considered in the counting of states. The main
quantitative result of our work is the correction of the 
value of the BH entropy.

%%%%%\section{Isolated horizons}

The standard definition of a BH as a spacetime region of no escape is a global definition. This notion of BH requires
a complete knowledge of a spacetime geometry and is therefore not
suitable for { describing local physics. 
%The physically relevant
%definition used, for instance, when one claims there is a ``black
%hole'' in the center of the galaxy, must be local.} One such local
%definition is the notion of isolated horizons (IH) introduced in
%\cite{Ashtekar:1999wa}.
{This is solved by using instead the notion of  Isolated horizons (IH): defined by extracting from the definition
of a Killing horizon the minimum conditions necessary for the laws
of BH mechanics to hold \cite{Ashtekar:1999wa}. They may be
thought of as ``apparent horizons in equilibrium''. Even though IH
are very general, allowing rotation and distortion, for simplicity
here we concentrate on the case in which the horizon
geometry is spherically symmetric. In the vacuum case, the latter are easy to 
visualize in terms of the characteristic formulation
%
% Just note of Jon to himself: the `characteristic formulation of GR' is the
% formulation where spacetime is foliated by null cones, as I found via google.
%
of general relativity with initial data given on null surfaces:
%
% I put the hyphen between spherical and IH to emphasize that spherical describes the IH
% and not the space-time --- important to convey the full generality of the case we
% are considering.
%
spacetimes with such IH} are solutions to Einstein's equations where
Schwarzschild data are given on the horizon and suitable free
radiation is given at a transversal null surface \cite{lewa} (see
Fig. \ref{figui}).

The calculation of black hole entropy in LQG is done by quantizing
the {\em sector} of the phase space of general relativity corresponding to
solutions having an IH. At the technical
level this {\em sector} is defined by
postulating the existence of a null boundary $\Delta \subset \sM$
with topology $S^2\times \R$ with { the pull-back of the
gravitational field to $\Delta$ satisfying the isolated horizon
boundary conditions.}

It is well known that the initial value formulation of general
relativity can be characterized in terms of a triad field
$e^i_a$ through $\Sigma=e\wedge e$---encoding the intrinsic spatial metric of $M$ as
$q_{ab}=e^i_ae^j_b\delta_{ij}$---and certain components of the
extrinsic curvature $K_{ab}$ of $M$ defined by $K_a^i=K_{ab}e^b_i$.
It can be shown  that the symplectic
structure of gravity
\begin{equation}
\label{sylstr1}
\Omega_{M}(\delta_1,\delta_2)=\frac{1}{8
\pi G}\int_{M} [\delta_1 \Sigma^{i} \wedge \delta_2
K^{\va}_{i}-\delta_2 \Sigma^{i} \wedge \delta_1 K^{\va}_{i}]
\end{equation}
is preserved in the presence of an IH. More precisely in the shaded
space-time region in Fig. \ref{figui} one has \be
\Omega_{M_2}(\delta_1,\delta_2)= \Omega_{M_1}(\delta_1,\delta_2).\ee
That is, the symplectic flux across the isolated horizon $\Delta$
vanishes due to the isolated horizon boundary condition
\cite{Ashtekar:1999wa, nouny}. One also has that, on shell, phase space
tangent vectors $\delta_\alpha, \delta_{v}$ of the form
\begin{eqnarray}\nonumber
\delta_{\alpha} \Sigma=[\alpha,\Sigma],\ \
\delta_{\alpha}K=[\alpha,K]; \ \ \delta_{v} \Sigma= \Lie_v \Sigma, \
\ \delta_{v}K= \Lie_v K
\end{eqnarray}
{for $\alpha: M \rightarrow \frak{su}(2)$ and $v \in \Vect(M)$
tangent to the horizon}, are degenerate directions of $\Omega_{M}$
from which one concludes that $SU(2)$ triad rotations and
diffeomorphisms are gauge symmetries \cite{cov}. Hence, the IH boundary condition { breaks neither the
symmetry under these diffeomorphisms nor the $SU(2)$ internal gauge
symmetry introduced by the use of triad variables.}
% \footnote{The IH
%boundary condition can be defined in terms of ($SU(2)$ gauge
%invariant) metric variables.}.
\begin{figure}[h]
\centerline{\hspace{0.5cm} \(
\begin{array}{c}
\includegraphics[height=4cm]{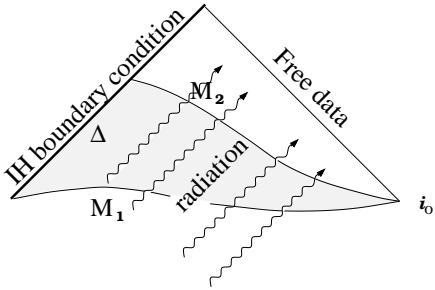}
\end{array}\) } \caption{The characteristic data for a (vacuum) spherically symmetric
isolated horizon corresponds to Schwarzschild data on $\Delta$, and free radiation data
on the transversal null surface.}
\label{figui}
\end{figure}

Ashtekar-Barbero connection variables are necessary for the quantization
{\em \`a la} LQG. When there is no boundary the $SU(2)$ connection \be
A_a^i=\Gamma^i+ {\beta} K_a^i\ee is canonically conjugate to
$\epsilon^{abc}\beta^{-1} \Sigma_{bc}^i/2$ where $\beta$ is the
so-called Immirzi parameter. As shown below, in the presence of a boundary the
situation is more subtle: the symplectic structure acquires a
boundary term $\Omega_H$. Due to the fact that, at the horizon, phase space
tangent vectors $\delta$ are linear combinations of $SU(2)$ gauge
transformations and diffeomorphisms tangent to the horizon $H=M\cap \Delta$, the
boundary term $\Omega_H$ is completely fixed by the requirement of gauge
invariance, i.e., the condition that local $SU(2)$ transformations,
now taking the form \be  \delta_{\alpha}\Sigma=[\alpha,\Sigma],\ \ \
\delta_{\alpha} A=-d_A \alpha\ee as well as diffeomorphisms
preserving $H$, continue to be degenerate directions of the
symplectic structure. The symplectic structure, in the new variables, becomes
\footnote{This follows firstly from $\Omega_M(\delta_{\alpha},\delta)=0$ implying
\begin{eqnarray*}&&
-\kappa \Omega_{\va H}\!=\! \int_{\va M}\!\!\!  \delta_{\alpha}
\Sigma_i \wedge \delta A^i\! - \! \delta \Sigma_i \wedge
\delta_{\alpha} A^i\!=\!
 \int_{\va M}\!\!\! [\alpha,\Sigma]_i \wedge \delta A^i \!+ \!\delta \Sigma_i \wedge d_A \alpha^i
\\ &&
\!=\!\int_{\va M}\!\!\! d (\alpha_i \delta\Sigma^i)\!-\!\alpha_i
\delta(d_A \Sigma^i)\! =\! - \frac{a_{\va H}}{\pi (1-\beta^2)}
\int_{\va H}\!\!\! \alpha_i \delta F^i(A) \\ && \!=\!
 \frac{a_{\va
H}}{\pi (1-\beta^2)}\!\!\! \int_{\va H} \!\!\!\delta_{\alpha} A_i
\wedge \delta
 A^i.
\end{eqnarray*} where we used the Gauss law
$\delta (d_A\Sigma)=0$,  condition (\ref{ihbc}), and that boundary
terms at infinity vanish. A similar calculation for diffeos
completes the proof (see \cite{nouny} for all details). } 
\ba
\label{BIsylstr} && \!\!\!\!\!\!\!\!\!\!\!\!\!\kappa \Omega_{M}=\int_{M}\!\!\! 2\delta_{[1}
\Sigma^{i} \wedge \delta_{2]} A^{\va}_{i}- {\frac{a_{\va H}}{\pi
({1-\beta^2})}} \int_H \!\!\! \delta_{1} A_i \wedge \delta_2 A^i, \ea
where $\kappa={8\pi G \beta}$, $a_{\va H}$ is the horizon area,
and we have used the IH boundary condition which in terms of
Ashtekar-Barbero variables is found \cite{nouny} to take the form \be
\label{ihbc} \Sigma^i+\frac{a_{\va H}}{\pi (1-\beta^2)} F^i(A)=0. \ee Note
that the boundary contribution to the symplectic structure is given by
an $SU(2)$ CS symplectic form. One { can also show directly}
that the boundary term contribution is necessary for time evolution to
preserve the symplectic form.

Another consequence of the fact that $SU(2)$ transformations and diffeomorphisms preserving $H$ are gauge is that
(in the canonical formulation) they are Hamiltonian vector fields generated by first class constraints. More precisely one has that
\ba
&& \n \Omega(\delta_{\alpha}, \delta)+\delta G[\alpha, A, \Sigma]=0,\\
&& \Omega(\delta_{v}, \delta)+\delta V[v, A, \Sigma]=0,\label{hamil}
\ea
where $G$ and $V$ are the Gauss and Diffeo constraints respectively. They take the form
\ba
&& \n \!\!\!\!\!\!\!\!\!\!\!\!\!\!\!\!\!G[\alpha, A, \Sigma]=\int_M  \alpha_i (d_A \Sigma^i/(\kappa\beta))\\ &&\ \ \ \ \ + \int_{H} \alpha_i \left[\frac{a_{\va H}}{\pi \kappa\beta (1-\beta^2)} F^i+ \frac{1}{\kappa\beta}{\Sigma^i}\right]\approx 0\n, 
%\\
\ea
for all $\alpha: M\to \frak{su}(2)$, and
\ba
&& \n \!\!\!\!\!\!\!\!\!V[v, A, \Sigma]=\int_M  \frac{1}{\kappa\beta}\left[\Sigma_i\wedge v\inter F^i -v\inter A_i d_A\Sigma^i \right] \\ &&\ \ \ \ \ \ \ \  - \int_{H} v\inter A_i \left[\frac{a_{\va H}}{\pi \kappa\beta (1-\beta^2)} F^i+ \frac{1}{\kappa\beta}{\Sigma^i}\right]\approx 0\n,
\ea
for all $v\in {\rm Vect}(M)$ that is tangent to $H$ at the horizon. Notice that the previous constraints have the usual  Gauss and diffeo constraint bulk-terms, plus boundary-terms given by smearings of (\ref{ihbc}) on $H$. Their Poisson algebra is 
 \ba\n && \{G[\alpha,  A, \Sigma],G[\beta, A, \Sigma]\}=
G([\alpha,\beta], A, \Sigma)\\ \n && 
 \{G[\alpha,  A, \Sigma],V[v, A, \Sigma]\}=G(\sL_v \alpha, A, \Sigma) \\ &&
  \{V[v,  A, \Sigma],V[w, A, \Sigma]\}=V([v,w], A, \Sigma),
 \ea
 where we have ignored the Poisson brackets involving the scalar constraint as its smearing must vanish on $H$, i.e.,  it does not affect the first class nature of the previous constraints. Thus  (\ref{ihbc}) 
 are first class constraints and can be implemented {\em \`a la} Dirac  in the quantum theory.
%
% Wasn't sure if I should add this:
%
% and using in part gauge-fixed identities
% in \cite{Ashtekar:1999wa}.}

%%%%%\section{Quantization}

{This fact and the form of the symplectic structure motivates one to handle the
quantization of the bulk and horizon degrees of freedom (d.o.f.)
separately.
%spaces, $\sH^{\va H}\otimes\sH^{\va B}$.  The
%physical Hilbert space is determined by imposing the IH
%boundary condition (\ref{ihbc}) as a quantum operator equation on
%$\sH^{\va H}\otimes\sH^{\va B}$ and then factoring out diffeomorphisms
%and imposing the bulk usual constraints\cite{bhe}.
%We first discuss the bulk quantization. 
As in standard LQG [8] one
first considers (bulk) Hilbert spaces $\sH^B_\gamma$ defined on a
graph $\gamma \subset M$  with end points on $H$, denoted $\gamma\cap H$.
The quantum operator associated with $\Sigma$ in (\ref{ihbc}) is
%
% As long as the other numerical factors are right, then this 4\pi G \gamma
% factor is what gives the matching of J^i(v) in the bulk with E_I^i in the surface
% Hilbert space, as is needed to have a viable matching of eigenvalues.
%
% *If* the correct factor here is *not* 4\pi G but something else, then
% the descrepancy may be in the relative normalizations of J^i and E_I^i
% as $su(2)$ valued objects. (Normalizations can most easily be defined
% by specifying the scaling of the structure constants, for example.)
% Another possible source of such a potential descrepancy is that a factor of 2
% was simply missed in one of the derivations.  But as of yet there is no descrepancy.
%
\begin{equation}
\label{gammasigma} \epsilon^{ab}\hat{\Sigma}^i_{ab}(x) = 8\pi G
\beta \sum_{p \in \gamma\cap H} \delta(x,x_p) \hat{J}^i(p)
\end{equation}
where $[\hat{J}^i(p),\hat{J}^j(p)]=\epsilon^{ij}_{\ \ k} \hat{J}^k(p)$ at each $p\in\gamma\cap H$.
%
% and $\epsilon^{ab}$ is the density weight 1 Levi-Civita tensor on $S$
%
Consider a basis of $\sH^{{\va B}}_{\gamma}$ of eigen-states of both
$J_p\cdot J_p$ as well as $J_p^3$ for all $p\in \gamma\cap H$ with
eigenvalues $\hbar^2 j_p(j_p+1)$ and $\hbar m_p$ respectively. These
states are spin network states, here denoted $|\{j_p,m_p\}_{\va
1}^{\va n}; {\van \cdots} \rangle$, where $j_p$ and $m_p$ are the
spins and magnetic numbers labeling $n$ edges puncturing the horizon
at points $x_p$ (other labels are left implicit). They are also
eigenstates of the horizon area operator $\hat a_{\va H}$
\[ \hat a_{\va H}|\{j_p,m_p\}_{\va 1}^{\va
n}; {\van \cdots} \rangle=8\pi\beta \ell_p^2 \,
\sum_{p=1}^{n}\sqrt{j_p(j_p+1)} |\{j_p,m_p\}_{\va 1}^{\va
n}; {\van \cdots} \rangle. \]
We can decompose $\sH^B_\gamma$
according to 
\begin{equation}
\sH_\gamma^B = \underset{\{j_p\}_{p \in \gamma \cap H}}{\bigoplus} \sH_\gamma^{\va B}({\{j_p\}})
%\otimes \left(\underset{p \in \gamma \cap H}{\otimes}{j_p}\right)
\end{equation}
%
%\begin{equation}
%\sH_\gamma^B = \underset{\{j_p\}_{p \in \gamma \cap H}}{\bigoplus} \sH_{\{j_p\}}
%\otimes \left(\underset{p \in \gamma \cap H}{\otimes}{j_p}\right)
%\end{equation}
for spaces $\sH_\gamma^{\va B}({\{j_p\}})$ spanned by states $|\{j_p,m_p\}_{\va
1}^{\va n}; {\van \cdots} \rangle$  for a given n-tuple $\{j_p\}$.

Substituting the expression (\ref{gammasigma})
 into (\ref{ihbc}) we get
\begin{equation}\label{seven}
-\frac{a_H}{\pi (1-\beta^2)}\epsilon^{ab}\hat{F}^i_{ab} = 8\pi G
\beta \sum_{p \in \gamma\cap H} \delta(x,x_p) \hat{J}^i(p)
\end{equation}
This equation tells us that the surface Hilbert space, $\sH^{{\va
H}}_{\gamma\cap H}$  is precisely the one
corresponding to (the well studied \cite{witten}) $SU(2)$ CS theory in the presence of {
particles}  with CS level $k=a_{\va H}/(2\pi\beta(1-\beta^2)\ell_p^2)$. 
The curvature of the (quantum) CS connection vanishes
everywhere on $H$ except at the position of the defects where
we find conical singularities of strength encoded in the quantum operators $\hat J^i_p$.

The solutions of (\ref{seven}) restricted to the graph $\gamma$ are found to be elements of the Hilbert space \cite{nouny}
\begin{equation} \sH_\gamma = \underset{\{j_p\}_{p \in \gamma \cap
H}}{\bigoplus} \sH_\gamma^{inv}({\{j_p\}}) \otimes \sH^{\va CS}_k({\{j_p\}}),
\end{equation}
where $\sH_\gamma^{inv}({\{j_p\}})$ is a proper subspace of $\sH_\gamma^{\va B}({\{j_p\}})$ spanned by area eigenstates, and
$\sH^{\va CS}_k({\{j_p\}})$ are the CS Hilbert spaces which turn out to be completely determined by the total spin of punctures $\{j_p\}$ \cite{witten}. 
%as the full kinematical Hilbert space for \mbox{fixed $\gamma$}.
The full Hilbert space  of solutions of
(\ref{seven}) is obtained as the projective limit of the spaces
$\sH_\gamma$.
%
%\footnote{To define the
%projective limit, one needs to define a family of consistent embeddings
%$\iota_{\gamma' \gamma} : \sH_{\gamma}^{kin} \rightarrow \sH_{\gamma'}^{kin}$
%whenever each edge of $\gamma$ can be obtained from edges of $\gamma'$ by
%composition and reversal of orientation.  One can use the usual embeddings
%appropriate for LQG --- namely pull-back via the projection $p_{\gamma'\gamma}$
%defined, e.g., in \cite{aldiffproj, alrev}; one can check that these embeddings
%preserve the $SU(2)$-invariance of the states, as well as the horizon invariance condition above,
%and therefore are also well-defined on the spaces $\sH_\gamma^{kin}$, and satisfy the
%necessary consistency conditions.}
%Finally, in order to obtain the physical Hilbert space it remains
%only to impose the diffeomorphism, and Hamiltonian constraints.
%The iffeomorphism constraints are imposed in the same way as in \cite{Ashtekar:1999wa}.
%The resulting Hilbert space $\sH^{Diff}$ is spanned by states which we may
%represent as $|n; \{j_p,m_p\}_{\va 1}^{\va n}; {\van \cdots} \rangle$.  Here $n$ is the
%number of punctures at the horizon, and $\{j_p,m_p\}_{\va 1}^{\va n}$
%is an ordered set of pairs $\{j_p,m_p\}$, one for each puncture; these are the data from the
%horizon Hilbert space that remain.  $\dots$ represent the rest of the data specifying the
%bulk state (which will not be relevant for this discussion).
The IH boundary condition implies that lapse must be zero at the
horizon so that the scalar constraint is only imposed in the
bulk.
%Because
%there is more horizon data in this case, this assumption is slightly
%stronger than in \cite{Ashtekar:1999wa}. }

%\section{State counting}
The entropy of the IH is computed by the formula $S={\rm tr}(\rho_{\va IH}\log\rho_{\va IH})$ where the density matrix
$\rho_{\va IH}$ is obtained by tracing over the bulk d.o.f., while restricting to horizon states that are compatible
with the macroscopic area parameter $a_{\va H}$. Assuming that there exist at least
one solution of the bulk constraints for every admissible state on the boundary, the entropy is given by
$S=\log(\sN)$ where $\sN$ is the number of  horizon states compatible with the given
macroscopic horizon area $a_H$. After a moment of reflection one sees that
\be {\sN}=\sum_{n;(j)_{\va 1}^{\va
n}}
%{\rm Sym}[n;\{j\}_{\va 1}^{\va n}]\
{\rm dim}[{\sH}^{\va CS}_k(j_1 {\van \cdots} j_n)], \ee where the
labels $j_1\cdots j_p$ of the punctures are constrained by the condition
\be \label{conki}a_H-\epsilon \le 8 \pi\beta
\ell_p^2 \, \sum_{p=1}^{n}\sqrt{j_p(j_p+1)}\le a_H+\epsilon . \ee
%
%The symmetry factor---given by \be {\rm
%Sym}[n;\{j\}_{\va 1}^{\va n}]\equiv \frac{n!}{\prod_j n_j!},
%\ee where $n_j$ denotes the number of punctures labelled by the spin
%$j$, { so that} $n=\sum_j n_j$---comes from the correct
%implementation of diffeomorphism invariance on the horizon
%\cite{bhe}.
Similar  formulae were first used in \cite{majundar}.

It turns out  that due to (\ref{conki}) we can compute the entropy for
$a_{\va H}>>\beta\ell_p^2$ (not necessarily infinite).  The reason
is that the representation theory of $U_q(SU(2))$---describing
${\sH}^{\va CS}_k$ for finite $k$---implies \be{\rm dim}[{\sH}_k^{\va
CS}(j_1 {\van \cdots} j_n)]={\rm dim}[{\rm Inv}(\otimes_p j_p)],\ee
as long as all the $j_p$ as well as the interwining internal spins
are less than $k/2$. But for
Immirzi parameter in the range $|\beta|\le\sqrt{3}$ this is
precisely granted by (\ref{conki}) \cite{nouny} . All this simplifies the entropy
formula considerably. The previous dimension corresponds to the
number of independent states one has if one models the black hole by
a single $SU(2)$ intertwiner!

Let us conclude with a few remarks.

We have shown that the spherically symmetric isolated horizon is
described by a symplectic form $\Omega_M$ that, when written in the
(connection) variables suitable for quantization, acquires a horizon
contribution corresponding to an SU(2) CS theory. Our
derivation of the (conserved) symplectic structure is
straightforward. We first observe that $SU(2)$ and diffeomorphism
gauge invariance is not broken by the IH boundary condition: they
continue to be degenerate directions of $\Omega_M$ on shell. This by
itself is then sufficient for deriving the boundary term that arises
when writing the symplectic structure in terms of Ashtekar-Barbero
connection variables (see also \cite{nouny}).

Note that no d.o.f. is available at the horizon in the classical
theory as the IH boundary condition completely fixes the geometry at
$\Delta$ (the IH condition allows a single (characteristic) initial
data once $a_{\va H}$ is fixed (see fig. \ref{figui})).
Nevertheless, non trivial d.o.f. arise as {\em would be gauge}
d.o.f. upon quantization. These are described by $SU(2)$
CS theory with (an arbitrary number of) defects which
couple to gauge d.o.f. through the dimensionless parameter
$16\pi^2\beta (1-\beta^2) \ell_p^2 \sqrt{j(j+1)}/a_{\va H}$, i.e.,
the ratio of a basic quantum of area carried by the defect to the
total area of the horizon. These {\em would be gauge} excitations
are entirely responsible for the entropy.

We obtain a remarkably simple formula for the horizon entropy: the
number of states of the horizon is simply given in terms of the (well
studied) dimension of the Hilbert spaces of CS theory with
punctures labeled by spins, which---due to the area constraint
(\ref{conki}), and for the range $|\beta|\le \sqrt{3}$ including the
physical value of $\beta$ \cite{barberos}---is just the
dimension of the singlet component in the tensor product of the
representations carried by punctures. The black hole density matrix
$\rho_{\va IH}$ is the identity on ${\rm Inv}(\otimes_p j_p)$ for
admissible $j_p$. Similar counting formulae have been proposed in the
literature \cite{models} by means of heuristic arguments. Our
derivation from first principles in particular clarifies previous
proposals.

General arguments and simple estimates
indicate that  the entropy will turn
out to be $S_{BH}={\beta_0} a_H/({4 \beta \ell^2_p})$,  where
$\beta_0$ is a constant to be determined.  The new counting
techniques of \cite{barba} are expected to be very useful for this.
Thus the result to leading order remains unchanged. However,
subleading corrections will have the form $\Delta S=-\frac{3}{2}
\log a_H$ (instead of $\Delta S=-\frac{1}{2} \log a_H$ 
in the $U(1)$ treatment) matching other approaches
\cite{amit}. This is due to the full $SU(2)$ nature of the IH
quantum constraints imposed here, { and
%---due to the universal nature of the logarithmic correction---
this is a clear-cut indication 
that the $U(1)$ treatment overcounts states}.  The value $\beta_0$ and the log-correction has been recently computed  for $|\beta| < \sqrt{3}$ \cite{barberos}.
 { The range $|\beta| \ge \sqrt{3}$ is unphysical as the quantum group structure imposes additional 
constraints driving the entropy below the physical value $a_{\va H}/4$.}

{ In ref. \cite{Ashtekar:1999wa}  the classical description of the IH was first done in terms of the null tetrad
formalism. In this case the null surface defining the Horizon provides the natural structure 
for a partial gauge fixing from the internal gauge  $SL(2,\C)$   to $U(1)$.  In this setting one fixes an internal direction $r^i\in \frak{su}(2)$ 
and the IH boundary condition (\ref{ihbc}) becomes 
%\vskip.1cm  \vskip-.1cm 
\ba
dV+\frac{2\pi}{a_{\va H}} \Sigma^i r_i=0, \ \ \ \Sigma^ix_i=0,\ \ \ \Sigma^iy_i=0, 
\ea
where $x^i,y^i\in \frak{su}(2)$  are arbitrary vectors completing an internal triad. In the quantum theory \cite{bhe} only the first of the previous constraints is imposed strongly, while---due to the non-commutativity of $\Sigma^i$ in LQG---the other two can only be imposed weakly, namely in \cite{bhe} one has
$\langle \Sigma^ix_i\rangle=\langle \Sigma^iy_i\rangle=0$. However, this leads to a larger set of admissible states (over counting).  
To solve this problem, within the $U(1)$ model, one would have  to solve the two constraints  $\Sigma^i x_i=0=\Sigma^iy_i$ at the classical level first, implementing the reduction
also on the pull back of two forms $\Sigma^i$ on $H$.  However, this would introduce formidable complications for the 
quantization of the bulk degrees of freedom in terms of LQG techniques. 
Our  $SU(2)$ treatment resolves this problem as now the three components of  (\ref{ihbc}) are first class constraints. 
Dirac implementation leads to a smaller subset of admissible surface states that are relevant in the entropy calculation. }

%\section*{Acknowledgements}
We thank A. Ashtekar, M. Knecht, M. Montesinos, D. Pranzetti, M. Reisenberger, and
C. Rovelli for discussions, and an anonymous referee for exchanges that have considerably improved the presentation of our results. This work was supported in part by the
Agence Nationale de la Recherche; grant ANR-06-BLAN-0050. J.E. was
supported by NSF grant OISE-0601844, and thanks Thomas Thiemann and
the Albert-Einstein-Institut for hospitality. A.P. is a member {\em
l'Institut Universitaire de France}.

\end{document}